\documentclass[iop,apj,numberedappendix,appendixfloats]{emulateapj}

\newcommand\pasa{PASA}
\graphicspath{{./}{figures/}}
\received{January 1, 2018}
\revised{January 7, 2018}
\accepted{\today}
\usepackage{ulem}
\DeclareGraphicsExtensions{.eps,.ps}

\shorttitle{Probing the emission states of PSR J1107$-$5907}
\shortauthors{Jingbo Wang et al.}

\begin{document}

\title{Probing the emission states of PSR J1107$-$5907}

\email{wangjingbo@xao.ac.cn}

\author{Jingbo Wang$^{1,2,3*}$, George Hobbs$^{4,9}$,  Matthew Kerr$^{5}$,  Ryan Shannon$^{6,7}$, Shi Dai$^{4}$, Vikram Ravi$^{8}$, Andrew Cameron$^{4,9}$, Jane F. Kaczmarek$^{4}$,  Robert Hollow$^{4}$, Di Li$^{9,10}$, Lei Zhang$^{9,4}$,  Chenchen Miao$^{9}$, Mao Yuan$^{9,}$, Shen Wang$^{9,11}$, Songbo Zhang$^{12,4}$, Heng Xu$^{13,14,9}$, Renxin Xu$^{13,14}$}
 \affiliation{$^1$ Xinjiang Astronomical Observatory, Chinese Academy of Sciences, 150 Science 1-Street, Urumqi, Xinjiang 830011, China }
\affiliation{$^2$ Key Laboratory of Radio Astronomy, Chinese Academy of Sciences, 150 Science 1-Street, Urumqi, Xinjiang, 830011, China}
\affiliation{$^3$ Xinjiang Key Laboratory of Radio Astrophysics, 150 Science1-Street, Urumqi, Xinjiang, 830011, China}
\affiliation{$^4$ CSIRO Astronomy and Space Science, PO Box 76, Epping, NSW 1710, Australia}
\affiliation{$^5$ Naval Research Laboratory, 4555 Overlook Ave., SW, Washington, DC 20375, USA}
\affiliation{$^6$ Centre for Astrophysics and Supercomputing, Swinburne University of Technology, P.O. Box 218, Hawthorn, Victoria 3122, Australia}
\affiliation{$^7$ ARC Centre of Excellence for Gravitational Wave Discovery (OzGrav)}
\affiliation{$^8$ Cahill Center for Astronomy and Astrophysics, MC 249-17, California Institute of Technology, Pasadena, CA 91125, USA}
\affiliation{$^9$ CAS Key Laboratory of FAST, National Astronomical Observatories, Chinese Academy of Sciences, Beijing 100012, China}
\affiliation{$^{10}$ NAOC-UKZN Computational Astrophysics Centre, University of KwaZulu-Natal, Durban 4000, South Africa}
\affiliation{$^{11}$ Astronomy Department, Cornell University, Ithaca, NY 14853, USA}
\affiliation{$^{12}$ Purple Mountain Observatory, Chinese Academy of Sciences, Nanjing 210008, China}
\affiliation{$^{13}$ Kavli Institute for Astronomy and Astrophysics, Peking University, Beijing 100871}
\affiliation{$^{14}$ Department of Astronomy, School of Physics, Peking University, Beijing 100871, China}

\begin{abstract}

	The emission from PSR~J1107$-$5907 is erratic. Sometimes the radio pulse is undetectable, at other times the pulsed emission is weak, and for short durations the emission can be very bright.  In order to improve our understanding of these state changes, we have identified archival data sets from the Parkes radio telescope in which the bright emission is present, and find that the emission never switches from the bright state to the weak state, but instead always transitions to the ``off'' state.  Previous work had suggested the identification of the ``off'' state as an extreme manifestation of the weak state.  However, the connection between the ``off'' and bright emission reported here suggests that the emission can be interpreted as undergoing only two emission states: a  ``bursting'' state consisting of both bright pulses and nulls as well as the weak-emission state. 
 
\end{abstract}


\section{Introduction} \label{sec:intro}

PSR J1107$-$5907 is an isolated radio pulsar that was discovered in the Parkes 20-cm Multibeam Pulsar Survey of the Galactic plane (\citealt{2013MNRAS.434..347L}).  Its rotational period (P $\sim$ 0.25\,s) is typical of normal pulsars, but a comparatively low period derivative ($\dot{P} \sim 9 \times 10^{-18}$).  places this pulsar in an under-populated region in the $P$-$\dot{P}$ diagram between the populations of normal and recycled pulsars. The inferred characteristic age of the pulsar ($\tau_c \sim$ 447\,Myr) indicates that it is amongst the oldest  non-recycled pulsars.

Pulsar emission is known to be complex.  Individual pulses vary in shape, phase and intensity.  At least one-third of pulsars exhibit the phenomena known as subpulse drifting (\citealt{2006A&A...445..243W}) in which individual pulse components drift in pulse phase. Many pulsars also exhibit ``nulling" (\citealt{1970Natur.228...42B}) during which the pulsed emission seemingly switches off for a few pulses at a time.  Pulsars are termed ``intermittent" if their emission ceases for long periods corresponding to a large number of missing pulses (for instance, the emission may cease for hours to years)\footnote{The term ``intermittent''  is not well defined. Some authors only use this term for pulsars whose emission ceases for timescales measured in days. Other authors use the term also for pulsars whose emission ceases on a much longer timescale than the pulse period, for instance over minutes to hours.}. \cite{2006Sci...312..549K}, \cite{Camilo12} and \cite{Lyne17} have shown that the spin-down rate for intermittent pulsars decreases when the pulse emission is off.   Other pulsars show discrete emission states in which the emission does not completely switch off in either state.  This is known as ``mode changing".  Various authors (e.g., \citealt{2007MNRAS.377.1383W} and \citealt{2010Sci...329..408L}) have suggested that mode changing and nulling are related phenomena. Pulsars with complex combinations of these  emission phenomena have also been reported (e.g., \citealt{2010MNRAS.408..407B} and \citealt{2019ApJ...877...55Z}).

Some pulsars, including PSRs~J1752+2359 (\citealt{2004ApJ...600..905L}), J1938+2213 (\citealt{2013MNRAS.434..347L}) and B0611+22 (\citealt{2014MNRAS.439.3951S}) exhibit ``bursting" emission during which a large number of bright individual pulses are detected over a relatively short time interval.  There is still no clear physical model for why this occurs. 



\begin{table*}
\centering
\caption{Observations that contain the bright emission state for PSR J1107$-$5907}\label{tb:observations}
\begin{tabular}{llllllllll}
\hline
\# &Filename                &  MJD            & Project ID& Receiver& Central Obs. freq& Bandwidth& Backend & Duration & Obs. mode\\
        &      &       &        &             & (MHz)    &    (MHz) &  & (s)       &\\
\hline
1&p120811\_231320.rf      &   56150.9  &  P574  &  MULTI  &   1382  &    400    &  CASPSR &  2712   &  fold \\
2&s121018\_213312.rf      &   56218.8  &  P832  &  MULTI  &   1369 &    256    &  PDFB3  &  21479 &  fold \\
3&s121019\_182134.rf      &   56219.7  &  P456  &  MULTI  &   1369 &    256   &  PDFB3  &  17879 &  fold \\
4&s140714\_095334\_3.sf &  56852.4  &  P863  &  1050CM &   732   &    64       &  PDFB3  &  900   & search \\
5&t140714\_095333\_3.sf &  56852.4  &  P863  &  1050CM &   3094   &    1024    &  PDFB3  &  900   & search \\
6&t141023\_222022.sf   &   56953.9  &  P863  &  1050CM &   3094 &    1024   &  PDFB4  &  417 & search \\
7&s141023\_222023.sf   &   56953.9  &  P863  &  1050CM &   732  &    64     &  PDFB3  &  417 & search \\
8&t160510\_124308\_1.sf &  57518.5  &  P863  &  1050CM &   3094 &    1024   &  PDFB4  &  900 & search \\
10&p160911\_021816.rf      &   57642.0  &  P863  &  H-OH   &   1382 &    400    &  CASPSR &   3616 &  fold \\
11&t180123\_02181.sf      &   58141.6  &  P863  &  MULTI   &   1369 &    256    &  PDFB4 &   900 &  search \\
12&t180603\_040025.rf      &   58272.1  &  PX500 &  MULTI  &   1369 &    256  &  PDFB4  &  7199 &  fold \\
\hline
\end{tabular}
\end{table*}

As we argue in this work, PSR~J1107$-$5907 exhibits a number of these emission features, including bursts.  \
\cite{2006ChJAS...6b...4O} reported that the pulsed emission of PSR~J1107$-$5907 switches between a null state with no detectable integrated pulse profile, a weak mode with a narrow pulse, and a bright mode, with a very broad pulse profile. In contrast, \cite{2014MNRAS.442.2519Y} argued that the pulsar only exhibits two emission states, strong and weak, with the previously proposed ``off'' state simply being an extreme end of the weak emission state. They also showed that the pulsar is likely a near-aligned rotator and that it does not exhibit any measurable spin-down rate variation between the emission states.  \cite{2016MNRAS.456.3948H} detected the pulsar using ASKAP (with some contemporaneous Parkes observations) and found  that the typical time-scale between the strong emission states is $\sim$3.7 hours and that the duration of the bright state was typically a few minutes, but in one case lasted for almost 40\,minutes. \cite{2018ApJ...869..134M} presented the first low-frequency detection of the pulsar with the Murchison Widefield Array (MWA) at 154 MHz and the simultaneous detection from the upgraded Molonglo Observatory Synthesis Telescope (UTMOST) at 835 MHz. They found that the pulsar exhibits steep spectral indices for both the bright main pulse component and the precursor component and the bright state pulse energy distribution is best parameterised by a log-normal distribution at both frequencies.


During the past few years this pulsar has been semi-regularly observed using the 64-m Parkes telescope. Most of these observations  are short, quick-look observations carried out as part of the P595 PULSE@Parkes outreach project (e.g., \citealt{2009PASA...26..468H}), interspersed with relatively long ($\sim$\,1 hour) observations obtained for a project studying pulsar intermittency (with Parkes observing code P863) and long rise-to-set observations carried out during time assigned to and supported by the Commensal Radio Astronomy FAST Survey (CRAFTS) program (\citealt{2018IMMag..19..112L}; with observing code PX500). The data sets comprise a number of observing modes and frequency bands and include both data folded at the known pulse period and ``search mode'' data in which single pulses can be analyzed.  

With an aim of assessing PSR~J1107$-$5907 identification as a bursting pulsar, we concentrate on the Parkes data sets that exhibit the strong bright state.  
In Section 2, we describe the observations and processing method. Our primary results are presented in Section 3 in which we study the occurrences of emission-state switches and study the single pulse energetics to support our hypothesis. We compare models of the emission states in Section 4 and conclude in Section 5.


\section{Observations}

The observations described here were obtained with the Parkes 64-m radio telescope. All these observations are archived in the CSIRO data archive (\citealt{2011PASA...28..202H}; data.csiro.au). Most are now publically available, but those acquired within the last 18 months are embargoed.  We visually inspected all the available observations of PSR~J1107$-$5907 and selected data files in which the pulsar was detected in its bright state, exhibiting a very broad and unambiguously bright integrated profile.  Table~\ref{tb:observations} lists, for each observation, the corresponding file name, the modified Julian date (MJD) of the observation start, project ID, observing frequency, observing mode (fold or search), backend instrument(s) used and the observation length.

Some observations were obtained simultaneously in the 10 and 40\,cm observing bands with the dual-band receiver, but the majority of the observations were in the 20\,cm observing band using the central beam of the 13-beam multibeam receiver.  A few were recorded using the H-OH single pixel receiver.  A number of backend systems have been used for recording the data, including the Parkes Digital Filterbanks (PDFB3 and PDFB4) and the CASPER Parkes Swinburne Recorder (CASPSR). Detailed descriptions of the receivers and backend systems can be found in \cite{2013PASA...30...17M} and \cite{2014MNRAS.442.2519Y}. 
The PDFB and CASPSR backend systems can be calibrated (both flux density and polarisation calibration) if (as is usually the case) a switched calibration signal was recorded either prior to or after the observation.  The final flux calibration is carried out by relating the calibration signal to the known flux density of Hydra A using the {\sc psrchive} pulsar signal processing system (\citealt{2004PASA...21..302H}, \citealt{2013PASA...30...17M} and \citealt{2019RAA....19..103X}).
We formed analytic templates for the bright and weak emission modes separately from our observations using {\sc paas} and  then obtained the flux density estimate using {\sc psrflux}, which matches the template with the observation and determines the area under the template.

 Search-mode data is folded to the pulse period or 30 second sub-integrations using the DSPSR (van Straten \& Bailes 2011) software package. The search-mode calibration files are folded at the calibration pulse period (11.123 Hz). We mitigated aliased signals and narrowband radio-frequency interference (RFI) by excising channels within 5\% of the band edge and those with a level substantially above a median-smoothed bandpass, respectively. The {\sc Psrchive} program {\sc pac} is used to perform flux (based on observations of Hydra~A) and polarimetric calibration. 


\section{Results}

\begin{figure*}

\centering
\includegraphics[width=23cm,angle=0,trim={4cm 0cm 0 0}]{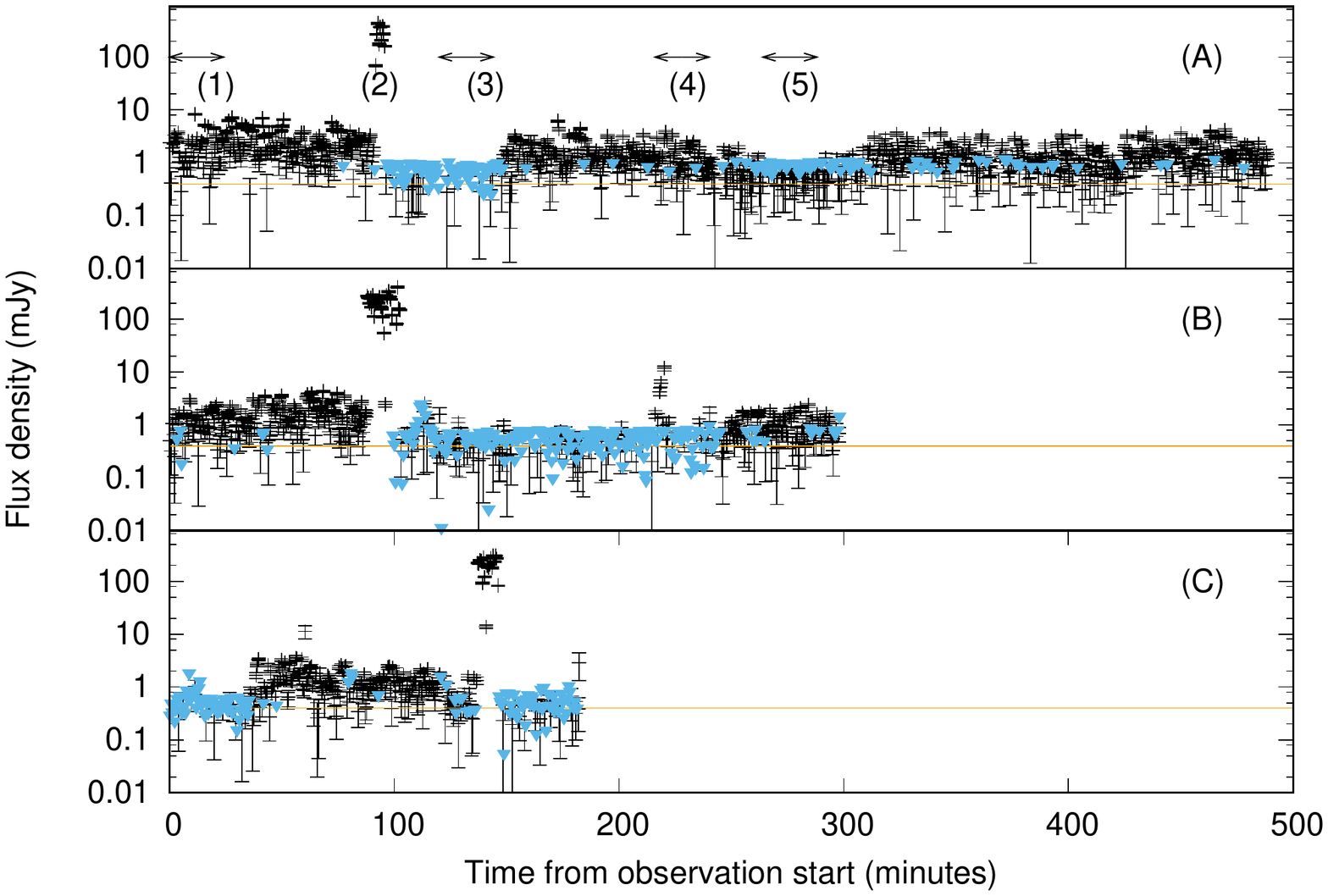}
\caption{The flux density variations of PSR~J1107$-$5907 for observations including the bright phase. The flux density is plotted on a logarithmic scale. The yellow line indicates a flux density of 0.4\,mJy. 3$\sigma$ upper bounds on the flux density are shown where no pulse is detectable as blue arrows.}\label{fg:fluxDensity}
\end{figure*}

In Figure~\ref{fg:fluxDensity} we show three representative examples of the flux density as a function of time for PSR~J1107$-$5907.  The three panels correspond to observations 12, 9 and 10 in Table~\ref{tb:observations} respectively. The short bright states have flux densities around $\sim$100\,mJy.  Note that the flux density is  plotted on a logarithmic scale and so where the measured flux density value is consistent with zero we plot an upper bound symbol (downward-pointing triangle) at the value of the 1$\sigma$ uncertainty.

\begin{table}
\caption{Occurrences of the various combinations of the possible state switches for the 12 observations that contain the bright state.}\label{tb:transitions}
\begin{center}
\begin{tabular}{ll}
\hline
Transition & Occurrences \\
\hline
weak to off           &  0\\
weak to bright        & 9 \\
bright to off         & 12 \\
bright to weak        & 0\\
off to weak           & 4\\
off to bright         & 3  \\ 
\hline
off to bright to off & 3\\
off to bright to weak & 0 \\
weak to bright to off& 9\\
weak to bright to weak & 0 \\
\hline
\end{tabular}
\end{center}
\end{table}

\begin{figure}
\centering
\includegraphics[width=8cm,angle=270,trim={0cm 3.5cm 0 0}]{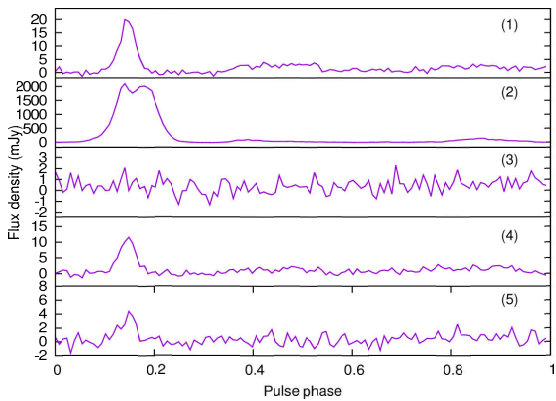}
\caption{The integrated pulse profiles from the five different observation segments marked in Figure 1.\label{fg:regions} }
\end{figure}

\begin{figure*}
\centering
\includegraphics[angle=270,width=18cm]{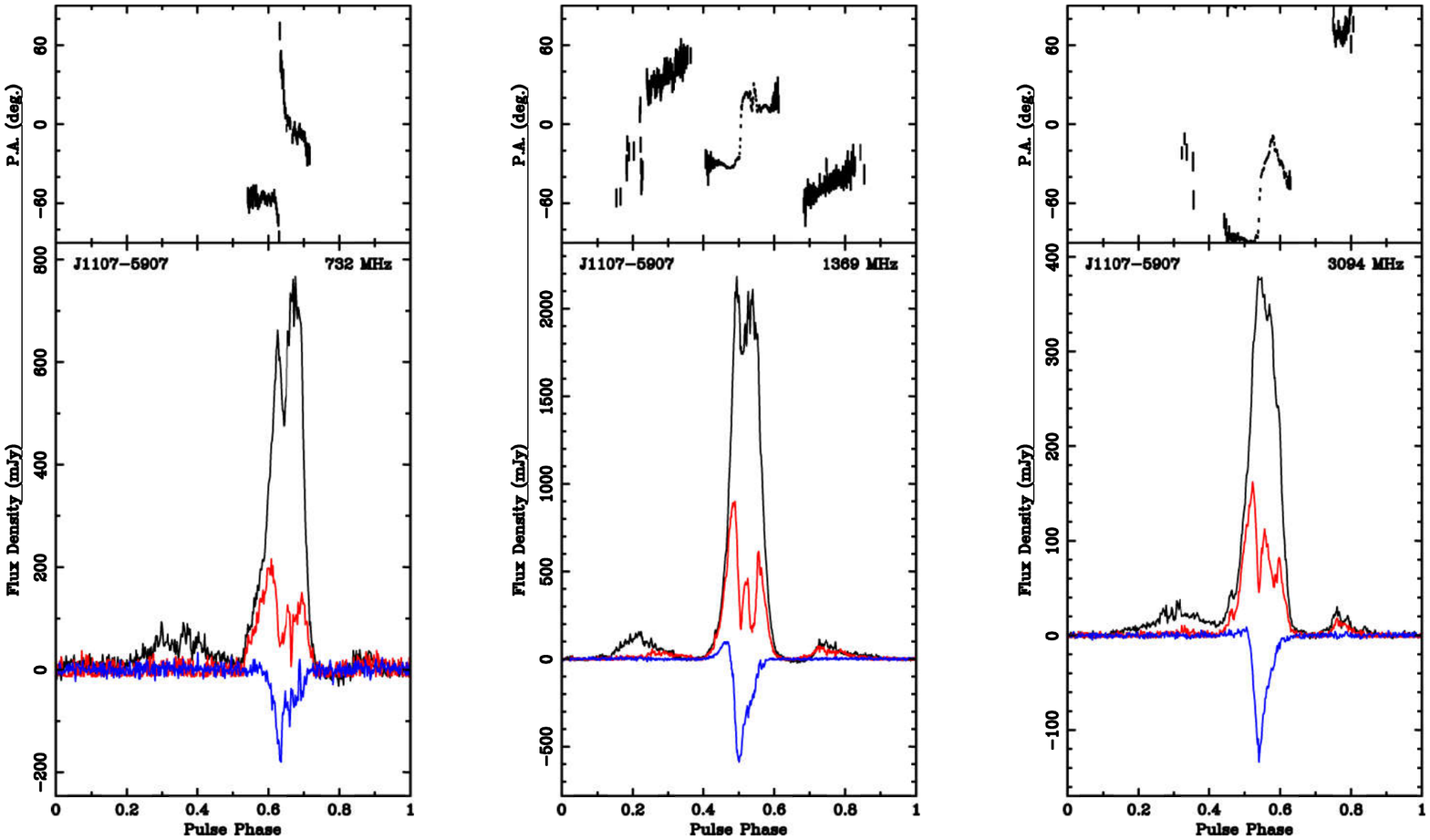}
\caption{ Averaged polarization pulse profiles centered during bright state at 10cm, 20 and 40 cm observing bands, respectively (from letf to right) . In each panel, the black line shows total intensity, red linear polarization, and blue circular polarization. The PA of the linear polarization is shown in the top panels. }\label{fg:pol}
\end{figure*}

For the longest observation (panel A in Figure~\ref{fg:fluxDensity}) we identify five regions (labelled 1 to 5).  All the regions apart from region 2 are of 24 minute duration. The folded, total intensity profiles
for these regions are shown in Figure~\ref{fg:regions}.  Note that the y-axes use different scales for each panel. The pulsar is much more polarized during the bright mode than in the weak mode and the linear and circular polarized profiles are similar in the three observing bands. As shown in Figure~\ref{fg:pol}, there are three main linear polarized components in the main pulse and the trailing component seems more linearly polarized than the precursor. As shown in  Figures~\ref{fg:fluxDensity} and \ref{fg:regions}, the mean flux density during the bright state is about two orders of magnitude higher than that of the weak state. We have checked through all the calibrated data sets and the mean flux densities in the weak state fluctuates within an order of magnitude in all of the three observing bands and ranges from 0.7 to 4.2\,mJy, 0.21  to 2.1\,mJy and 0.07  to 0.54\,mJy in the 10\,cm, 20\,cm and 50\,cm observing bands, respectively.


\begin{figure}
\centering

\includegraphics[width=9.5cm,trim={2cm 0cm 0 0}]{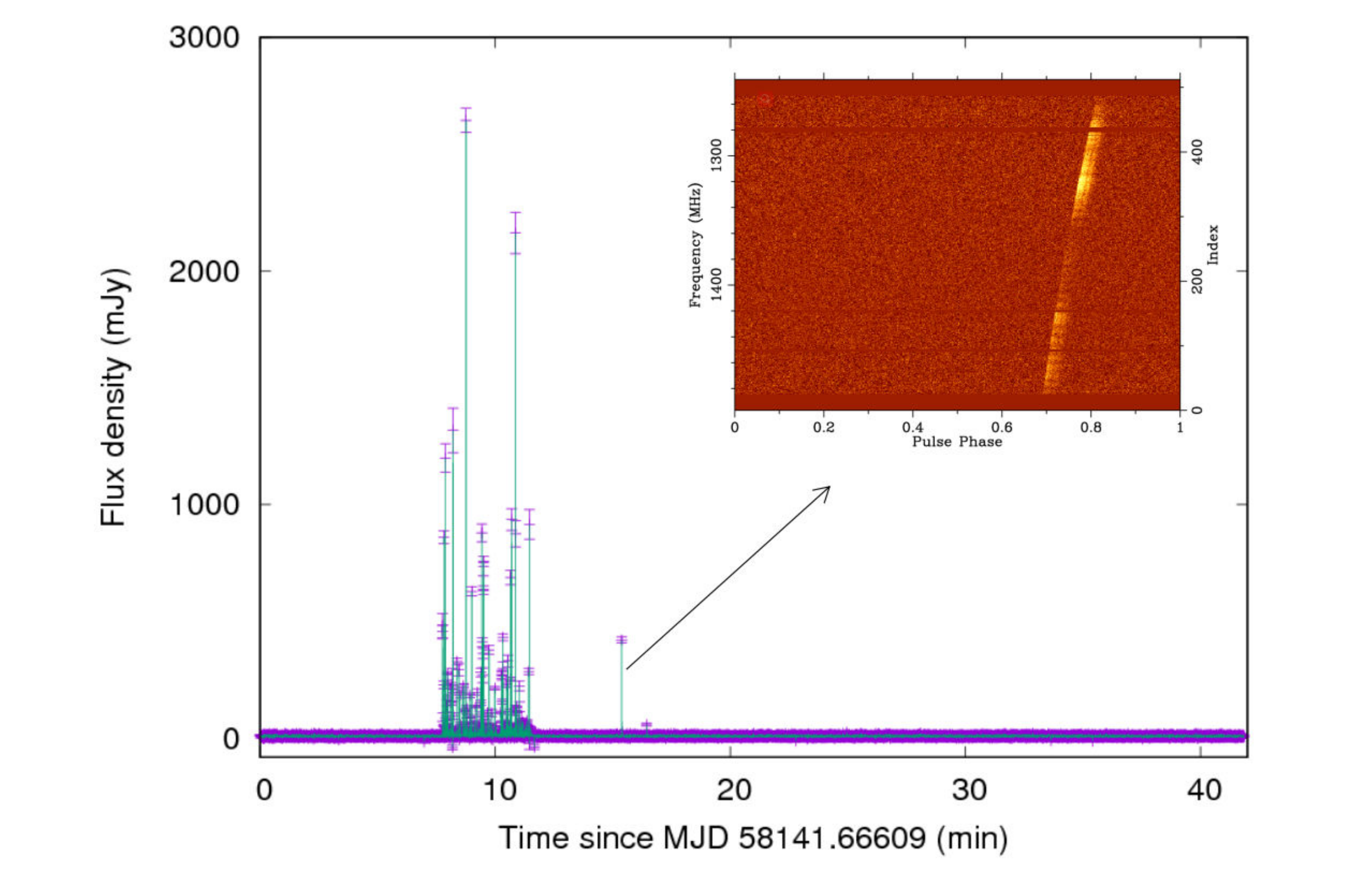}
\caption{Calibrated single pulses fluxes versus time during state transitions from ``off'' to
``bright" to ``off''.  The inset shows frequency versus phase for the strong single pulse that occurred after the bright state.}\label{fg:singlePulse}
\end{figure}

The pulsar flux density shown in Region (1) of Panel (A) of Figure~\ref{fg:fluxDensity} is around 2\,mJy at the start of the observation. This is typical of the weak state that has been described by \cite{2014MNRAS.442.2519Y} and \cite{2006ChJAS...6b...4O}.  There is a sudden change to the bright state (Region 2). The duration of the detected bright states ranges from a few minutes to a few tens of minutes and the intensity of strong single pulses during bright states can be higher than 10\,Jy (as shown in Figure~\ref{fg:singlePulse}) at 0.7\,GHz. All the integrated profiles during bright states are similar but not identical since the strength and the number of single pulses during each bright state are different.
After the bright state in panel (A) the pulsar seems to switch off.  In region (3) we have no significant measurements of a pulse with an upper limit of $< 0.2 $\,mJy.  The folded pulse profile (Figure~\ref{fg:regions}) shows no indication of a pulse.  We explain below (Section 4) that this is a likely an indication of a bursting-emission state, and not, as \cite{2014MNRAS.442.2519Y}  suggest, part of the weak emission state.
Region (4) is back to the weak state (although the mean flux density of 1.2\,mJy is slightly lower than the mean value prior to the bright emission).  Between regions (4) and 5 we identify a significant decrease in the flux density.  However, in contrast to Region (3), in Region (5) the emission is still present (as seen in the bottom panel of Figure~\ref{fg:regions}) and the mean flux density is 0.4 mJy.  This is the so-called low-level ``underlying emission'' identified by \cite{2014MNRAS.442.2519Y} which is only detected through profile integration and commonly seen in our data set.  The emission then increases back to a typical weak state level.

For this observation (panel A) the emission pattern is from weak at the start of the observation to bright, to off, and then to weak (though the final weak state varies significantly in flux density until the end of the observation).  For all of the observations listed in Table~\ref{tb:observations} we have determined whether the emission transitions from ``weak to off'', ``bright to off'', ``bright to weak'', ``off to weak'' or ``off to bright''.  These results are presented in Table~\ref{tb:transitions}.  In all observed cases the emission switches from ``bright to off'', and never from the bright to the weak state.  The majority of the observations exhibit the pattern ``weak to bright to off'', but in a few cases (observation numbers 7, 10 and 11 in Table~\ref{tb:observations}) we observe ``off to bright'' and then back to off. 

A more detailed look at an ``off to bright to off'' transition can be seen in Figure~\ref{fg:singlePulse}. In this Figure we have shown the flux density of individual  pulses at 1.4 GHz around a bright emission state.  For the first $\sim 7$\,minutes we see no evidence of any emission.  Then the pulsar abruptly enters the bright emission state. The bright state lasts for $\sim$5\,min and then the emission switches back off. However, we notice two individual, single pulses occurring just after 15 minutes from the start of the observation.  These pulses are clearly from the pulsar (the inset shows the dispersion of the pulse).  Based on the single pulse observations listed in Table~\ref{tb:observations}, we have determined that the pulsar seems always to enter the bright emission state suddenly, within one rotation of the neutron star.

\begin{figure}
\centering
\includegraphics[width=6cm,angle=-90]{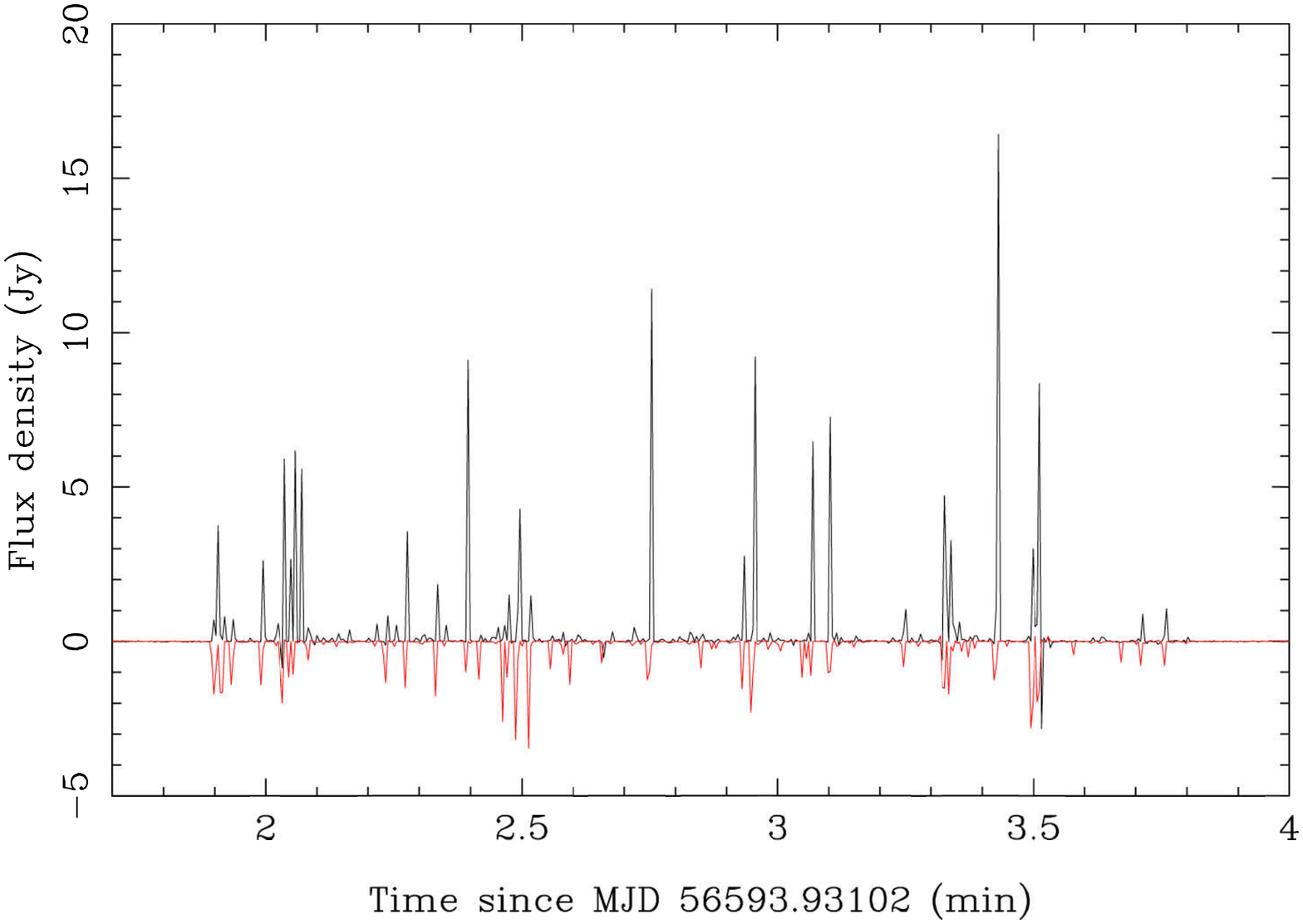}
\caption{The dual-band (10/40cm) single pulse flux density variation during a bright states. The 40\,cm data are shown as positive flux densities in black and the 10\,cm data as negative values in red.}\label{fg:dualband}
\end{figure}

We have two observations in which simultaneous dual-band (10/40 cm) single-pulse data sets were recorded whilst the pulsar was in the bright state (observation numbers 4, 5, 6 and 7 in Table~\ref{tb:observations}). The calibrated single pulse flux densities variation are shown in Figure~\ref{fg:dualband} for observation numbers 6 and 7 (note that observation files 6 and 7 represent DFB3 and DFB4 simultaneous observations of the pulsar in different observing bands). In order to compare the flux density in the two bands, the flux densities in the 10\,cm observing band are shown as negative values. The flux densities for the sporadic single pulses seen during the weak state are on a level of a few tens of mJy, one hundred mJy and a few hundreds of mJy in the 10, 20 and 50\,cm bands, respectively. They are more than an order of magnitude weaker than the strong single pulses in the bright state as shown in Figure~\ref{fg:singlePulse} and Figure~\ref{fg:dualband}.

These 453 single pulses shows that the emission abruptly starts and stops, but the emission is ``bursting''; the individual pulses vary significantly in terms of their flux densities and spectral index. We identify 28 pulses in which we clearly detect the pulse in the 40\,cm observing band, but not in the 10\,cm band and  40 pulses in which the pulse is only observed in the high-frequency band. We calculated the spectral index from 60 individually matched pulses (above a signal-to-noise threshold of 6 at both bands). The spectral index range from $-$3.56 to 1.37 with a mean of $-$1.11 and a standard deviation of 1.1. The variation  of the spectral index is more extreme than that presented by \cite{2018ApJ...869..134M}.

\begin{figure*}
\centering
\includegraphics[width=8.5cm]{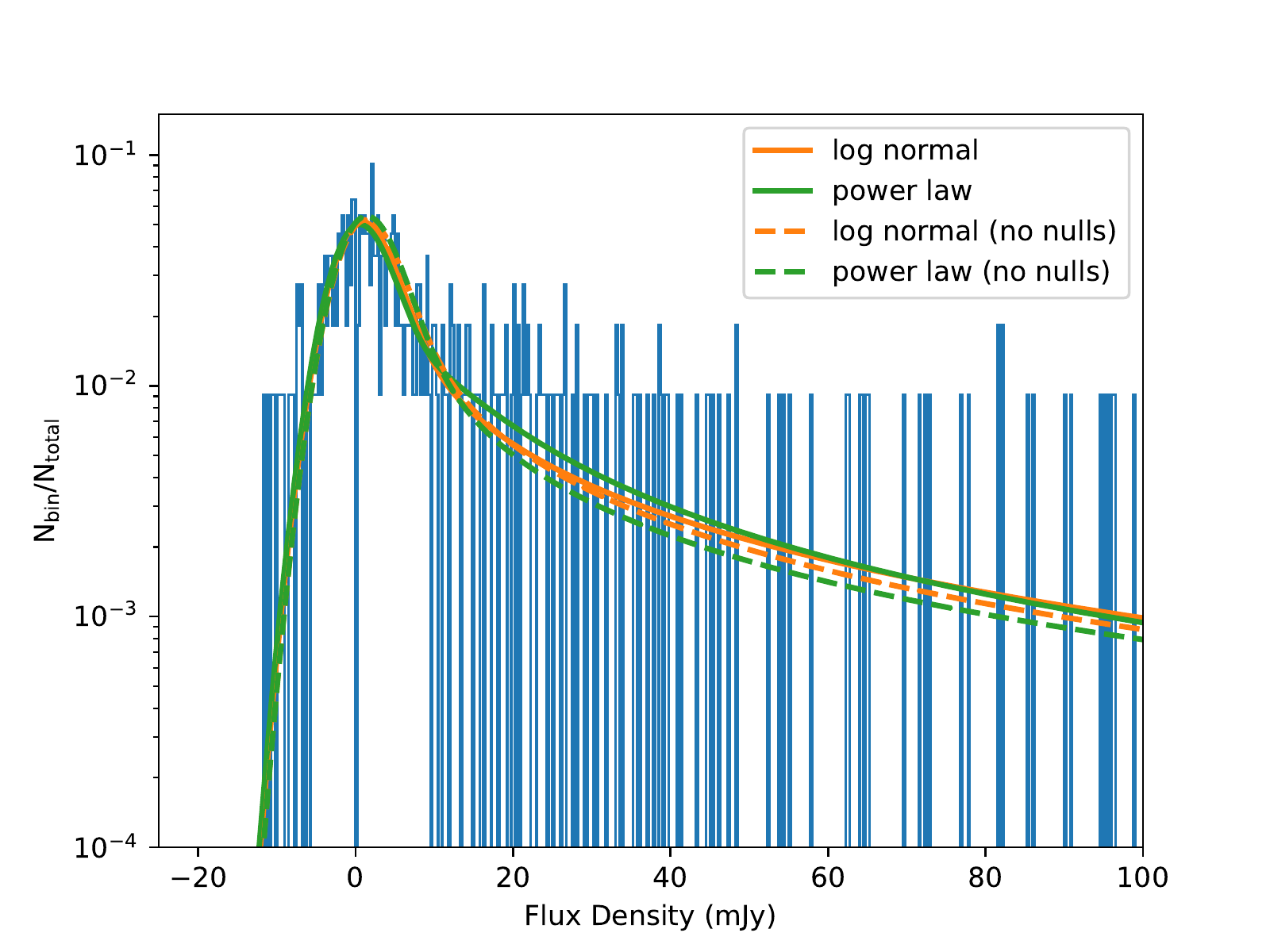}
\includegraphics[width=8.5cm]{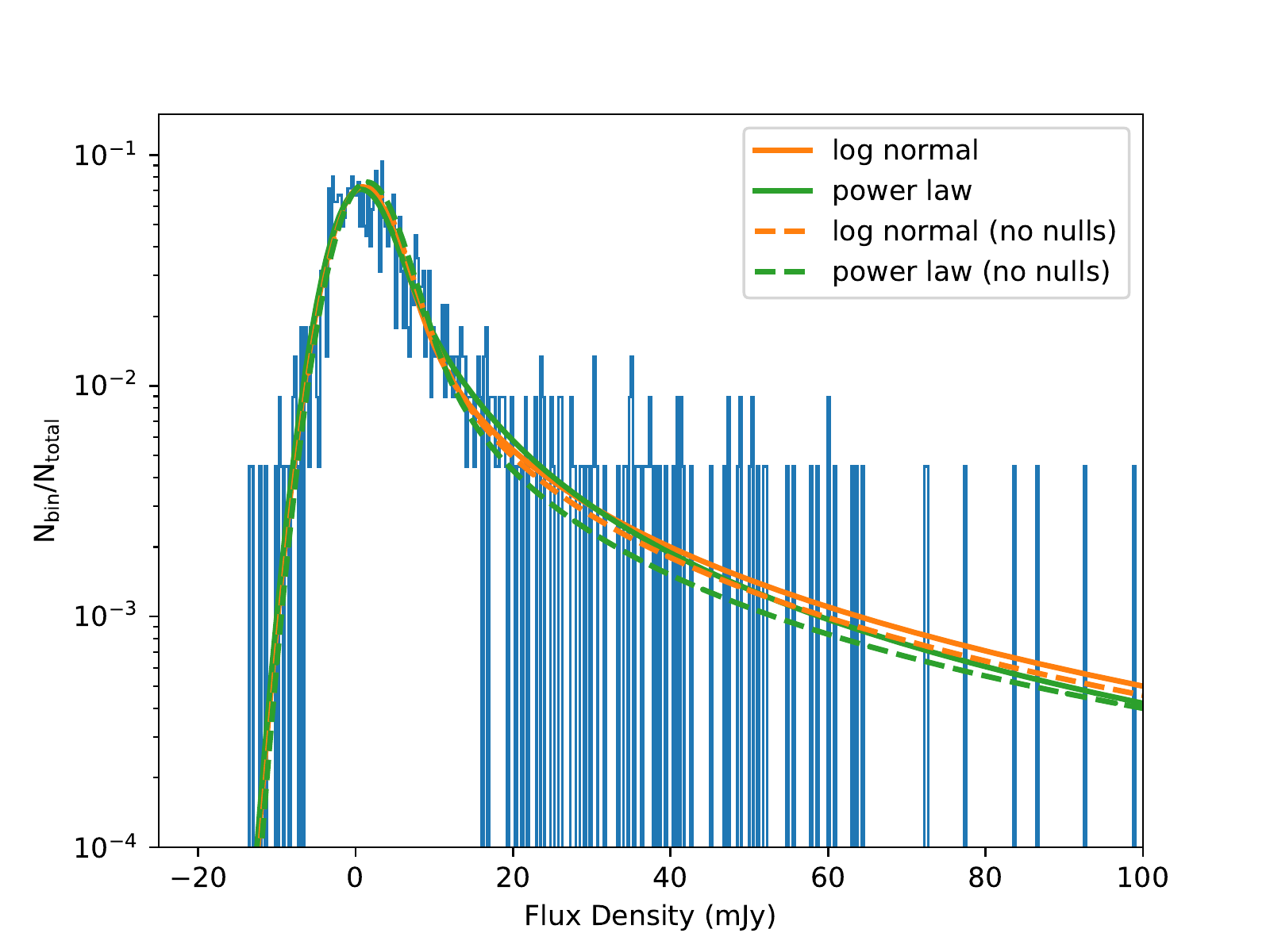}
\caption{The pulse energy distributions and best fits models during bright state for observation No.6 and No.11 (from left to right) in Table~\ref{tb:observations}.}\label{fg:pulseenergy}
\end{figure*}

Characterising the pulse energy distribution of a pulsar is helpful in understanding the pulse emission process. 
The pulse energy distribution of pulsars can often be represented by single component distributions (a log-normal or a power-law distribution). Since our data are well calibrated, we identify the pulse energy with the flux density. Below, we fit the pulse energy distribution for the bright states of observations No.~6 and No.~11 in Table~\ref{tb:observations} (see also Figure~\ref{fg:singlePulse} and the red line in Figure~\ref{fg:dualband}) and display the resulting best-fit model in Figure~\ref{fg:pulseenergy}.

We modelled the pulse energy distribution with both a log normal and a power law distribution, and we also considered an additional null component for each case.  In detail, we first characterized the white noise properties of the pulse distribution, finding it to be well-described by a normal distribution, and then convolved the trial distribution (log normal or power law) with the white noise model to determine the predicted distribution of pulse intensities.  We determined the parameters of the distribution by optimizing the gaussian likelihood. The specific form of the power law, $P(F) \propto (1+ (F/F_c)^2)^{\alpha/2}$ includes a low-energy cut-off, $F_c$ in addition to the spectral index $\alpha$.  The log normal distribution follows the standard form $P(F)\propto (1/\sigma F)\exp[-0.5\,(\ln F-\mu)^2/\sigma^2]$ with parameters for the logarithmic width ($\sigma$) and mean ($\mu$) of the distribution.  We ignored a small number of negative outliers which were formally below $4\sigma$ for our white noise distribution, as well as the long region of nulls shown in Figure~\ref{fg:singlePulse}. We incorporate nulls using a mixture model whose relative normalization (nulling fraction) is the only additional free parameter.

After optimizing the parameters for both models, we find the power law distribution is preferred by an increase in the log likelihood of $\sim$4 for each of the two data sets.  While this increase is formally significant, given the potential influence of systematic errors in the flux densities, we argue that both models provide an adequate description of the pulse energy distribution.  We also considered the presence of an additional component of nulls, interleaved with the pulse energies described by the power law and log normal distributions.  With this component, the log likelihood improves by $\sim$ 4 for the log normal model for both sets, and by 18 and 13 for the power law.  A null component is, formally, strongly preferred for the power law model, though we again caution that systematic flux errors could dilute this preference.  The best-fit nulling fraction for the log-normal model is 25\% and 30\% for the two data sets, and 35 and 45\% for the power law model, respectively.  (We again emphasize we have excluded the obvious long span of nulls for observation No.11 in Table~\ref{fg:singlePulse}.)  In summary, there is no strong preference for either model, and only modest evidence for nulling under the log-normal description.  However, if the power law model is the correct underlying distribution, there is good evidence for a nulling fraction of 35--45\%.

\section{Discussion}

We believe that there are three possible descriptions of the emission states in this pulsar:

\begin{enumerate}
    \item As originally described by O'Brien et al. (2006), the pulsar exhibits three emission states: weak, strong and off. 
    \item As proposed by \cite{2014MNRAS.442.2519Y}, there are only two emission states: weak and strong. Here, the ``off'' state is only the weakest part of the weak state.
    \item A new description in which there are only two states, which we label ``persistent" and ``bursting". The persistent state is the same as the ``weak" state described earlier. However, the ``bursting" state contains very bright single pulses (leading to the bright emission), but also long periods of nulls (leading to the off state).
\end{enumerate}

We note that more sensitive telescopes may be able to detect emission from the off state. 
Our current observational results are inconsistent with description \#2 as there seems to be a repeating pattern off ``bright" to ``off", which is hard to explain if the off state is simply a low-S/N part of the weak state.


Distinguishing between descriptions \#1 and \#3 is difficult. But a model with nulls occurring aside bright pulses is mildly preferred. We have tried to measure the pulse frequency and its derivative during different states. However, the measured precision of  these quantities is insufficient to distinguish them. The number of nulls between two bright single pulse in the bright state ranges from 1 to $\sim$ 1200 (as shown in Figure~\ref{fg:singlePulse}. The bright state often ends with many nulls as shown in Figure~\ref{fg:singlePulse}, but the number of nulls after the bright state varies widely between observations. 
During the weak state, many detectable single pulses can often be seen and these detectable single pulses form the visible integrated pulse profile. The flux density of these detectable single pulses is on the level of $\sim$ 100\,mJy at 1.4\,GHz and a few tens of mJy at 3.0\,GHz,  which is much weaker than single pulses that occur during the bright state. As shown in Figure~\ref{fg:fluxDensity}, the flux densities fluctuate rapidly over a small range during the weak state.  Scintillation structure is clearly visible in the single dispersed pulse from the  inset of Figure~\ref{fg:singlePulse}, raising the possibility that interstellar scintillation could influence the inferred properties of the single pulses.  We estimate a scintillation bandwidth of 73 MHz from a frequency-domain auto-correlation analysis.  It is more difficult to estimate the scintillation timescale since the pulsar is in weak state for the most of time, however, no obvious change in the scintillation structure is observed during the longest bright state, setting a lower limit on the diffractive timescale of 24 minutes.  We can thus conclude that the inferred properties of single pulse are not affected by interstellar scintillation. 

Pulsar state switching behaviors including nulling, bursting, mode changes, sub-pulse drifts, and long-term intermittency can all be modeled as Markov processes (\citealt{2013ApJ...775...47C}). \cite{2014MNRAS.445..320K} modelled the nulling of PSR J1717$-$4054 as a three-state Markov process. However, in order to model the state switching of the pulsar in this paper, the different states need to be well defined via the single pulse properties first. Quantitative evaluation of a Markov model also requires observations with sufficient S/N. 

 Even for many of the well-studied objects, pulse-to-
pulse variations cause overlap of on-and-off-state intensities,
leading to false positives from algorithms that identify state
changes. Future work with now-available wideband receiver
systems on existing telescopes and eventually with new array
telescopes (ASKAP, MeerKAT, and the SKA) can improve the
discrimination between states as well as expand the sample of
objects that can be studied in this way.





\section{Conclusion}
Our analysis of PSR~J1107$-$5907 observations is not consistent with the \cite{2014MNRAS.442.2519Y} model in which the pulsar exhibits only a bright and a weak emission state. However, our results are consistent with a two-state model in which the pulsar is in a persistent or in a bursting-emission state. 

Continued monitoring of this pulsar is essential to understand the emission in more detail, but there are two clear areas that could be improved from the current Parkes observations. First, a much more sensitive telescope could probe the pulse nulls/very weak/off state in more detail and then different states of the pulsar can be well defined and used as input for a Markov model of the state switching process.  PSR~J1107$-$5907 is in the far Southern sky and so, until the SKA-era, the only telescope capable of providing more sensitive observations is MeerKAT, which is about 7 times more sensitive of Parkes multi-beam observations.  Second, the dual-band 10/40\,cm data indicates significant variations in the single pulse properties across that band. The recently installed ultra-wide-bandwidth receiver at Parkes provides uninterrupted coverage over the entire band with greater sensitivity than currently possible.

\section{Acknowledgements}
The Parkes radio telescope is part of the Australia Telescope, which is funded by the Commonwealth of Australia for operation as a National Facility managed by the Commonwealth Scientific and Industrial Research Organisation (CSIRO). This paper includes archived data obtained through the CSIRO Data Access Portal (http://data.csiro.au). This work is supported by the Youth Innovation Promotion Association of Chinese Academy of Sciences, 201$^*$ Project of Xinjiang Uygur Autonomous Region of China for Flexibly Fetching in Upscale Talents, the National Key R\&D Program of China (No.2017YFA0402602), the National Natural Science Foundation of China (No.11690024) and the Strategic Priority Research Program (B) of the Chinese Academy of Sciences (No.XDB230102000). R.S. acknowledge Australian Research Council grant FL150100148.  Work at NRL is supported by NASA.


\end{document}